\documentclass[useAMS,usenatbib]{mn2e}  
\usepackage{graphicx}
\usepackage{txfonts}
\newcommand{\xmm}{{XMM-{\em Newton} }}
\newcommand{\xmmns}{{XMM-{\em Newton}}}
\newcommand{\swift}{{\em Swift }}
\newcommand{\swiftns}{{\em Swift}}
\newcommand{\chandra}{{\em Chandra }}
\newcommand{\rosat}{{\em ROSAT }}
\newcommand{\rosatns}{{\em ROSAT}}
\newcommand{\integral}{{\em INTEGRAL }}
\newcommand{\ngc}{{NGC~3599 }}
\newcommand{\ngcns}{{NGC~3599}}

\newcommand{\lumUnits}{{ergs s$^{-1}$ }}
\newcommand{\lumUnitsns}{{ergs s$^{-1}$}}
\newcommand{\unit}[1]{\ensuremath{\, \mathrm{#1}}}
\newcommand{\msolar}{{$\unit{M_{\odot}}$ }}
\newcommand{\msolarns}{{$\unit{M_{\odot}}$}}
   \title[The soft X-ray flare in NGC~3599]{Was the soft X-ray flare in NGC~3599 due to an AGN disc instability or a delayed tidal disruption event?}

\author[R.D. Saxton et al.]{R.D. Saxton$^{1}$\thanks{E-mail:
richard.saxton@sciops.esa.int}, 
S.E. Motta$^{1,2}$,
S. Komossa$^{3}$ and 
A.M. Read$^{4}$\\
$^{1}$ESAC, Apartado 78, 28691 Villanueva de la Ca\~{n}ada, Madrid, Spain\\
$^{2}$University of Oxford, Dept. of Physics, Denys Wilkinson building, Keble road, OX1 3RU, Oxford, U.K.\\
$^{3}$Max Planck Institut f\"ur Radioastronomie, Auf dem  H\"ugel 69, 53121 Bonn, Germany\\
$^{4}$Dept. of Physics and Astronomy, University of Leicester, Leicester LE1 7RH, U.K.}

\begin{document}
\date{Accepted 1988 December 15. Received 1988 December 14; in original form 1988 October 11}

\pagerange{\pageref{firstpage}--\pageref{lastpage}} \pubyear{2002}

   \maketitle
   %\offprints{R. Saxton}
\label{firstpage}

\begin{abstract}
We present unpublished data from a tidal disruption candidate in \ngc
which show that the galaxy was already X-ray bright 18 months before the
measurement which led to its classification. This removes the possibility that 
the flare was caused by a classical, fast-rising, short-peaked, 
tidal disruption event. 
Recent relativistic simulations indicate that the majority of disruptions 
will actually take months or years to rise to a peak, which will then 
be maintained for longer than previously thought. \ngc could be one of 
the first identified examples of such an event.
The optical spectra of \ngc indicate that it is a low-luminosity 
Seyfert/LINER with $L_{\rm{bol}}\sim10^{40}$ \lumUnitsns. The flare may 
alternatively be explained by a thermal instability in the accretion disc, 
which propagates through the inner region at the sound speed, causing
an increase of the disc scale height and local accretion rate.
This can explain the $\leq 9$ years rise time of the flare.
If this mechanism is correct then the flare may repeat on a timescale 
of several decades as the inner disc is emptied and refilled.
\end{abstract}


%
%________________________________________________________________

\section{Introduction}

Tidal disruption events \citep[TDE;][]{Hills,Luminet,Rees88}
occur when a stellar object is destroyed and subsequently accreted by a
super-massive black hole (SMBH). Ten to twenty events have been detected in the
optical \citep[e.g.][]{Komossa08,Wang11,vanVelzen11,Cenko12b,Gezari12,Arcavi14} 
%and UV (\cite{Gezari06}, \cite{Gezari08}, \cite{Gezari09}) bands, 
and UV \citep{Gezari06,Gezari08,Gezari09} bands, 
where they show a modest increase in luminosity
over the integrated galaxy emission, and a characteristic light curve which lasts
weeks to months. 
TDE are most spectacular in the X-ray band where the contrast 
between the low-level X-ray emission of the host galaxy and the huge accretion flare, can reach 
factors of 1000s \citep{Komossa1242,Halpern04,Komossa05}. 
TDE were first identified from soft X-ray flares
seen in optically quiescent galaxies by the \rosat mission \citep{Bade96,Komossa99b,KomossaBade,Grupe99,Greiner}.
A small number have also been detected in higher-energy X-rays by 
\swift \citep{Burrows11,Bloom,Cenko,Pasham}
and \integral \citep{Walter13} while new soft-X-ray TDE have been identified by 
comparing  \xmm slew survey data with earlier \rosat data \citep{Esquej07,Saxton12} and by other X-ray searches \citep{Cappelluti09,Maksym10,Lin11,Maksym14,KhabSaz14}.
The presence of a pre-existing accretion disc may enhance the tidal disruption rate \citep{KarasSubr07}
and \citet{Merloni} have proposed that up to 10\% of optically-selected AGN could be caused by stellar disruptions. On an individual level, large flares
in persistent AGN present a certain ambiguity. The classic case is IC~3599 where a giant X-ray and
UV flare, which repeated after 20 years, has been alternatively explained 
by tidal stripping of an orbiting star \citep{Campana15} or by an accretion 
disc instability \citep{Grupe15}.

\ngc ($\alpha_{2000}$ = $11^{\rm h} 15^{\rm m} 26.^{\rm s}9$, 
$\delta_{2000}$ = $+18^{\circ} 06' 37^{''}$, z=$0.0028$) was discovered in an 
\xmm slew from 2003 with a soft X-ray flux a factor $>100$ higher than an upper limit from \rosat
\citep{Esquej07}. Subsequent observations of the galaxy 
by \xmmns, \chandra and \swift revealed a strong decay in flux by a factor $\sim100$
over the following years \citep{Esquej08,Esquej12}.
The galaxy shows weak, narrow, optical lines leading to its classification
as a LINER or Seyfert 2 galaxy \citep{Esquej08}.

Three scenarios were proposed at the time to explain the flare: a TDE, AGN variability and an
ultra-luminous X-ray source (ULX; based on the relatively low X-ray luminosity, $L_{\rm{X}}\sim$ a few $\times10^{41}$ ergs/s).
In this paper we reassess these possibilities based on newly discovered high-state and more recent
low-state data. In particular we investigate whether the flare could be due to an accretion disc instability, similar to that proposed for flares seen in certain galactic accreting binaries \citep{Cannizzo96,Belloni97a}. We also look at 
recent advances in numerical and analytical modelling of TDE lightcurves
\citep{Guill15,Hyak15,Shiok15,Piran15}, which show a 
generally slower rise to peak flux than that predicted by the classical model \citep{Rees88}.

A $\Lambda$CDM cosmology with ($\Omega_{\rm{M}},\Omega_{\Lambda}$) = (0.27, 0.73)
and  $H_{0}$=70 km$^{-1}$s$^{-1}$ Mpc$^{-1}$ has been assumed throughout.

\section{The soft X-ray flare in \ngc}

\ngc was reported as a candidate tidal disruption event, based on an X-ray flare seen 
in an \xmm slew from 2003-11-22 \citep{Esquej07}. The source flux
was $\sim 150$ times higher than that seen in a \rosat pointed
observation from 1993 \citep{Esquej12}. 
The galaxy was sparsely monitored post-flare by \xmmns, \swift and Chandra 
revealing a decline in X-ray flux of a factor 100 from the peak 
value \citep{Esquej08,Esquej12} and a shape that can be reasonably fit
by a canonical $t^{-5/3}$ curve, appropriate for the rate of
return of tidal debris to the disruption radius \citep{Rees88,Phinney}. 
A Chandra observation from 2008 pinpointed the emission to be
within 60 pc of the nucleus 
%, which taken together with the scale of the
%variability and the spectral shape, effectively excluded the possibility that
%the flare could be from a ULX 
\citep{Esquej12}.

Optical spectra were taken in 2007 and 2008 and found to be consistent with a pre-flare spectrum \citep{Caldwell,Esquej08}. After subtraction of the galactic
contribution, weak, narrow, low-ionisation lines remain whose
ratios lie on the border between LINER and low-luminosity
AGN activity \citep{Kauffmann04,Esquej08}. 
The luminosity of the narrow [OIII]$\lambda$5007 line was $L_{\rm{OIII}}=1.1\times10^{38}$
\lumUnits implying a historical bolometric luminosity, $L_{\rm{bol}}^{\rm{hist}}\sim10^{40}$ \lumUnits applying the standard correction factor for low-luminosity AGN
\citep{Lamastra}. During the flare the soft X-ray luminosity was
$L_{0.2-2}=5.5\times10^{41}$ \lumUnitsns. No simultaneous data are
available from other wavelengths making the calculation of $L_{\rm{bol}}$
at peak quite uncertain. Based on the reasonable assumption that the soft X-ray
flux was dominated by thermal disc emission, as suggested by 
the very soft spectrum \citep{Esquej07}, then a large fraction of the 
peak bolometric luminosity may be emitted in the soft X-rays.
Even so this implies an increase in
$L_{\rm{bol}}>50$ from the historical to the peak luminosity. 

\begin{figure}
\centering
%\rotatebox{-90}{\includegraphics[height=8cm]{ngc3599_lc_2.ps}}
\rotatebox{-90}{\includegraphics[height=9cm]{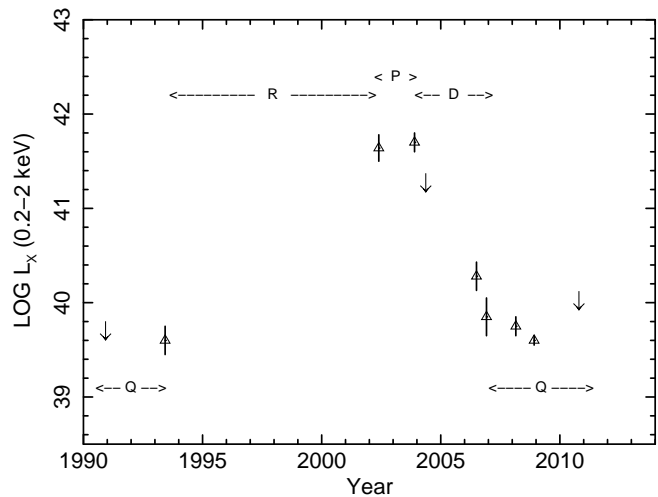}}
\caption[\ngc light curve]
{\label{fig:lcurve} The 0.2--2 keV X-ray luminosity light curve of \ngcns. 
Points are from \rosatns, the \xmm slew and \xmmns, \swiftns-XRT and \chandra
pointed observations. The last upper limit has been made by combining \swift 
data from 2010-10-23 and 2010-11-16. Flare phases are marked as: quiescent (Q),
rise (R; $\leq107$ months), plateau (P; $\geq18$ months) and decay 
(D; $\sim36$ months).} 

\end{figure}

Improvements to the \xmm science analysis software \citep[SAS;][]{Gabriel} have recently allowed some archival slew data to be processed for the first time. Amongst these was the slew 9045100003,
which passed over \ngc in 2002-05-22. Analysis of this
slew surprisingly reveals 19 photons from \ngc in 4.0 seconds of effective 
exposure time, yielding a soft X-ray luminosity, 
$L_{0.2-2}=4.8\pm{1.1}\times10^{41}$ \lumUnitsns, consistent 
with the luminosity seen at the peak in 2003-11-22. The source, then,
was already bright 18 months before the peak flux was
measured and more than a year before the date of the presumed disruption
itself, derived by fitting a canonical $t^{-5/3}$ curve to the X-ray 
measurements \citep{Esquej08}. The historical light curve is plotted in
Fig.~\ref{fig:lcurve} and includes the flux from 2002-05-22, a new upper 
limit from \xmm slew 9081400004,
taken on 2004-05-20, and an upper limit from a merge of 2 \swift observations
made on 2010-10-23 and 2010-11-16. We have analysed the \rosat observation of
1993-06-06 (rp600263n00) and identify \ngc with the catalogued source, 
1WGA~J1115.4+1807, with a count rate of $0.0026\pm{0.0006}$ c/s. 
This equates to a flux 130 times lower than that seen in 2002-05-22.

\begin{figure}
\centering
\rotatebox{-90}{\includegraphics[height=9cm]{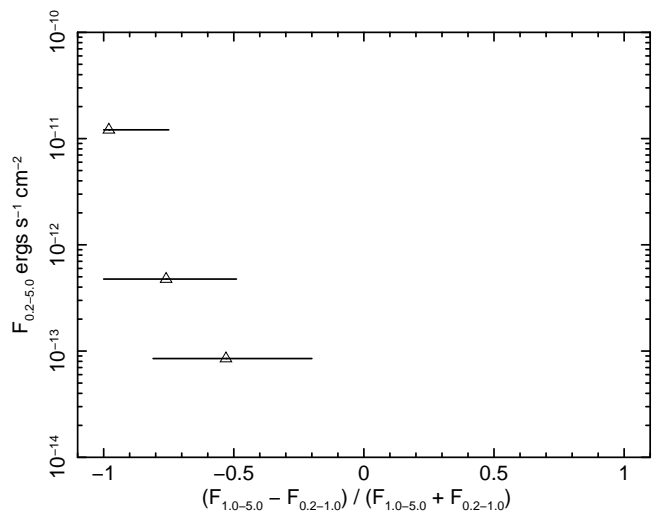}}
\caption[\ngc hardness v intensity plot]
{\label{fig:hplot} The hardness-intensity diagram for \ngcns, displayed
as the ratio of the unabsorbed 1.0--5.0 keV flux to the 0.2--1.0 keV flux 
plotted against the total unabsorbed 0.2--5.0 keV flux.}

\end{figure}

\citet{Esquej12} showed that the \xmm and \chandra observations all 
had soft spectra and could be simultaneously fit with a black-body 
of $kT\sim45$ eV plus a steep power-law, with $\Gamma\sim2.7$, both absorbed
by the Galactic column \citep[$1.42\times10^{20} $ cm$^{-2}$;][]{Kalberla}.
 We fit this model, with free black-body temperature and power-law slope, to each 
individual observation, in Fig~\ref{fig:hplot}, and display the hardness ratio as a function of
source flux. The hardness ratio is defined in terms of the 1--5 keV
and 0.2--1.0 keV fluxes as $H_{\rm{r}}$=($F_{1-5} - F_{0.2-1}$) / 
($F_{1-5} + F_{0.2-1}$) and remains soft even in the lowest flux states.
 
\section{Discussion}
\subsection{A Tidal Disruption Event}

The classical model of TDE predicts that, after the disruption, streams of debris will quickly
shock, lose energy and accumulate most of the bound material at a radius, $2R_{\rm{P}}$,
where $R_{\rm{P}}$ is the pericentre of the stellar orbit \citep{Rees88}. From here it will accrete
quickly causing a rapid rise in luminosity.
The TDE candidates detected to date, which have been well-monitored before the peak, have all shown a short rise time consistent with this model.
In NGC~5905,
$t_{\rm{rise}}$ was a few days \citep{Bade96}; in PTF10iya,
$t_{\rm{rise}}\leq0.03$ years \citep{Cenko12b}, in PS1-10jh, 
$t_{\rm{rise}}\sim0.21$ years \citep{Gezari12} and in three other Palomar Transient Factory (PTF) 
candidates from $\sim0.08 - 0.14$ years \citep{Arcavi14}.

The time when the fallback maintains a peak accretion rate is also
seen to be very short observationally \citep[e.g.][]{Bade96, Cenko12b, Gezari12, Arcavi14}. 
At face value, the event in \ngc plateaus between 2002-05-22 and 2003-11-22 and then
declines sharply at a rate compatible with $t^{-5/3}$. This behaviour is quite different from 
the light curves of previous, well-monitored, TDE. 
In principle, the event in \ngc could still be a tidal disruption event, similar
to the ones previously observed, if the 
2002-05-22 measurement lies on the rise of the luminosity curve and the 2003-11-22 measurement on the decline.
\citet{Esquej08} obtained a date for the onset of the monotonic decline of 
the luminosity curve, 
$t_{0}=2003.59\pm{0.06}$.
If we adopt a minimum start date of 2003.53 then the new slew data 
predate the decline from peak by 1.14 years. If this {\it were} a tidal disruption,
 then the
rise time, plus the time where the debris fall back rate is maintained
near maximum, is $\geq 1.14$ years; significantly longer than seen in 
previous TDE candidates.
This excludes the possibility that the flare in \ngc was
caused by a classical, {\em prompt} (fast-rising), TDE.

A slow, factor 5, flux decay in the TDE candidate, RBS~1032, was observed in ROSAT observations spanning 3.5 years \citep{Maksym14,KhabSaz14}. These measurements may represent a flat,
extended peak emission. They are, however, consistent
with a canonical $t^{-5/3}$ decline curve first detected $\sim 1$ year after peak \citep{Maksym14}.

%Although arguments have been made for longer peak times, 
%depending on the internal structure of the disrupted object (\cite{Lodato09}),
%this result together with the coincidence of the consistent luminosity in the
%2002 and 2003 observations 
%and the fairly modest (factor $\sim100$) increase in $L_{bol}$ 
%makes it now seem unlikely that \ngc was a TDE.

Recently, numerical simulations and new analytical work have shown that the
development of a TDE light curve is dependent on when the streams of tidal debris intersect 
each other \citep{Guill15,Hyak15,Shiok15}. Early interactions 
appear to be rare
and in the majority of cases circularisation occurs late and at a large distance from the BH, 5--10 times further away than predicted by the classical model
 \citep{Piran15}. This leads to a longer viscous timescale and a rise to peak that takes months, years or even decades \citep{Guill15},  especially when $M_{\rm{BH}}\lesssim6\times10^{6}$\msolarns.
 \citet{Esquej08} found a black hole mass for \ngc of 
$M_{\rm{BH}}=0.7-4.1 \times10^{6}$\msolar
from the $M_{\rm{BH}}-\sigma$ relation and 
$M_{\rm{BH}}=3.5\pm{0.9} \times10^{7}$\msolar from the relationship
with the absolute V magnitude of the galaxy given by \citet{Lauer07}.
A further measure of $M_{\rm{BH}}=3.1\pm{1.5} \times10^{7}$\msolar
is available from the K luminosity of the galaxy bulge \citep{Caramete10},
giving a full range of possible
black hole masses, $M_{\rm{BH}}=7\times10^{5}-4.6\times10^{7}$\msolar.
 
 The discussion above addresses the complete disruption of a main sequence star.
 Evolved stars have weakly-bound envelopes which may be
 stripped in black-hole encounters without destroying the star \citep{MacLeod12}. These events, which 
 result in light curves with a long rise to peak and a plateau phase lasting years, may be difficult to
 distinguish from a late circularisation after the disruption of a main sequence star. 

As \citet{Guill15} point out, the fact that we have only seen {\em prompt} TDE so far is likely
to be a selection effect as {\em delayed} (slow-rising, long-plateau) TDE, which achieve lower peak luminosities
than prompt TDE \citep{Shiok15}, being less spectacular, are more difficult to detect in transient surveys.  
This may be the reason why they have not been seen yet in optical surveys\footnote{Although they 
could be hiding in long baseline surveys such as Catalina \citep{catalina}.} or during the
\rosat mission. The comparison of \xmm and \rosat observations 
gives a baseline which
currently stretches to 25 years and is hence not biased against slow-rising TDE.
\citet{KhabSaz14} identified a small number of new TDE candidates by comparing
\rosat and \xmm pointed observation source fluxes. Of these, only RBS~1032 (discussed 
above) had sufficient ROSAT observations to be able to detect a potential 
delayed TDE
\footnote{ \xmm pointed observations currently cover just 2\% of the sky and the \xmm
slew survey is only deep enough to detect strong variability in the brightest sources.
Comparisons with future all-sky surveys such as eRosita \citep{eRosita} will undoubtedly increase the number of 
ROSAT-detected TDE.}. 
  \ngc could then be one of the first detections of a {\it delayed} tidal disruption event. 

\subsection{A ULX}
The peak luminosity of \ngcns, $L_{\rm{X}}=5\times10^{41}$ \lumUnitsns, falls
within the range attained by ULX \citep{HLX1}. 
Some arguments against a ULX interpretation for the 
flare in \ngc were presented in \citet{Esquej12}; we summarise these here and add new analysis.

A low-state \chandra observation revealed two X-ray sources within the galaxy. A brighter one,
coincident with the nucleus, with a 90\% confidence error of 60 pc, and a fainter source, with $L_{\rm{X}}=10^{39}$
\lumUnitsns, at a distance of 250 pc.  The \xmm pointed observation of 2006 had a flux 6 times higher
than the \chandra observation and is coincident with the nucleus with a 1-sigma error of 60 pc, 
excluding the possibility that the fainter \chandra source caused the variability.  The luminosity of
the \chandra nuclear source is consistent with that expected from the optical emission lines, hence
it is highly likely that this represents the low-level AGN in its normal historical state. This \chandra source
has an unusually steep spectrum, similar to that seen in the \xmm observations 
\citep[a power-law with $\Gamma=2.7$ plus a black body of kT=44 eV;][]{Esquej12}. 
Stellar-mass accreting binaries and ULX have significantly harder low-state 
spectra than this 
\citep[e.g. $\Gamma\sim1.4-2.1$ was found for the low-luminosity 
observations of HLX-1;][]{Servillat11}. For these reasons we locate the 
flare in the galactic nucleus rather than from a nearby ULX.

\subsection{A Highly Variable AGN}
In section 2 we saw that \ngc is a low-luminosity Seyfert/LINER and it is worth re-considering the possibility 
of an AGN flare in the light of the new data.
Any AGN variability mechanism needs to explain the following characteristics
of the flare in \ngcns: (i) an increase of factor 130 in X-ray flux 
within $\leq 9$ years; (ii) an unusually soft peak spectrum ($\Gamma\sim4$ /
kT$\sim90$ eV); (iii) a spectrum which remained soft while the
flux dropped by a factor 100 between 2003 and 2008.

%We start from an accretion disc model, where radiation is produced by thermal emission from a multi-temperature plasma and by the reprocessing of disc photons from a corona of hot electrons.
%We consider two main mechanisms which can cause large variability:  a change in line-of-sight absorption and a change in the structure of the disc.

No obvious intrinsic absorption was found in the low-state \chandra and \xmm spectra \citep{Esquej08,Esquej12}.
As a further test, 
%Closer material may be ionized, giving rise to the multitude 
%of complex spectral effects seen in warm absorbers (e.g. \cite{Blustin};\cite{KomossaFink97}). 
%Variable ionised absorption has been proposed to explain the factor 400,
%soft ($\Gamma=6$) X-ray flare seen in WPVS~007 (\cite{Grupe95WPVS}).
%This is supported by the discovery of variable mini-BAL outflows
%in UV spectra (\cite{Leighly09}) which prove the existence of high
%density, ionized material in the vicinity of WPVS~007. In this scheme the abnormally soft X-ray 
%spectrum is produced by an absorber which is transparent below the
%Oxygen VI edge (0.56 keV) and opaque to higher photon energies. 
we investigated the possibility that the decline in flux in \ngc is
due to a variable warm absorber by fitting the two pointed \xmm observations
from 2006 and 2008 simultaneously, with a fixed power-law model and an
ionized absorber (zxipcf) which was allowed to vary. 
The best fit was poor, with
$\chi^{2}_{\rm{r}}=127/61$, ruling out that the source variability is caused 
by a single phase absorbing medium.  Warm absorbers, when looked at 
with sufficient spectral resolution,
often appear to be multi-phased, containing gas at different distances and
ionisation states \citep[e.g.][]{Longinotti13}. The \ngc spectra have neither
the spectral resolution nor statistics to be able to exclude variable
multiple absorbers. However, such a scenario would not explain why the peak 
luminosity, during the \xmm slew observations, was $>50$ times higher than 
the historical bolometric luminosity nor why the spectrum remained 
soft while the flux reduced by a factor 100. From this we infer that 
the source experienced an increase in intrinsic emission rather than a variation
in a patchy absorber.

\subsubsection{A change in the disc structure}

A change in the distribution of matter, involving a filling and emptying of the inner accretion disc, has been invoked to explain long-lived changes in the emission state of solar-mass black-hole binaries 
 \citep[BHB;][]{Esin97}. 

The time $\tau_{\rm{fill}}$ taken for material to migrate from a truncation radius and fill the inner disc
up to the innermost stable circular orbit (ISCO) is governed by the viscous time scale and can be calculated from the mass of material in this inner region and the accretion rate, such that:

\begin{equation}
   \tau_{\rm{fill}} = M_{\rm{inner}} / \dot{M}
\end{equation}

The mass of the inner disc being given by: 

\begin{equation}
   M_{\rm{inner}}= \int_{R_{0}}^{R_{\rm{trunc}}}\rho(r)\,2\pi r\,H(r)\,dr
\end{equation}

where $R_{0}$ is the radius of the ISCO, $R_{\rm{trunc}}$ is the truncation radius, 
$\rho(r)$ is the disc density and H(r) its height.

For a Shakura-Sunyaev thin disc \citep{ShakSun73}, \citet{FKR} give the disc density and height as

\begin{equation}
    \rho(r) = 3.1\times10^{-8} \alpha^{-7/10}\dot{M}_{16}^{11/20}M_{1}^{5/8}R_{10}^{-15/8}f^{11/5} \unit{g\, cm^{-3}}
\end{equation}
\begin{equation}
    H(r) = 1.7\times10^{8} \alpha^{-1/10}\dot{M}_{16}^{3/20}M_{1}^{-3/8}R_{10}^{9/8}f^{3/5} \,\mathrm{cm}
\end{equation}

where $R_{10}$ is the disc radius in units of $10^{10}$ cm, $M_{1}$ is the mass of the black hole in solar
masses, $\dot{M}_{16}$ is the accretion rate in units of $10^{16}$ g$s^{-1}$ and $\alpha$ is the viscosity parameter.

Converted to AGN scaled units (with the radius expressed in gravitational radii, 
$R_{\rm{g}}=GM/c^{2}$, the accretion rate in units of the Eddington-limited accretion rate, $M_{\rm{edd}}=1.4\times10^{18} M_{1}$ g$s^{-1}$ and the black hole mass $M_{6}$
in units of $10^{6} M_{\odot}$, the inner disc mass is then given by:

\begin{equation}
    M_{\rm{inner}}= 6 \times 10^{-4}\alpha^{-8/10}M_{6}^{11/5}M_{\rm{edd}}^{-3/10} \Big[\big(\frac{R_{\rm{trunc}}}{R_{\rm{g}}}\big)^{5/4} - \big(\frac{R_{0}}{R_{\rm{g}}}\big)^{5/4}\Big] 
    \,\unit{M_{\odot}}
\end{equation}

and the filling time $\tau_{\rm{fill}}$ by:

\begin{equation}
    \tau_{fill} \sim 0.33\alpha^{-8/10}M_{6}^{6/5}M_{\rm{edd}}^{-3/10}\Big[\big(\frac{R_{\rm{trunc}}}{R_{\rm{g}}}\big)^{5/4} - \big(\frac{R_{0}}{R_{\rm{g}}}\big)^{5/4}\Big] \,\mathrm{months}
\end{equation}

Note that $\tau_{\rm{fill}}$ is equivalent to the viscous
timescale of a slim disc at the truncation radius.

\begin{figure}
\centering
\rotatebox{0}{\includegraphics[height=7cm]{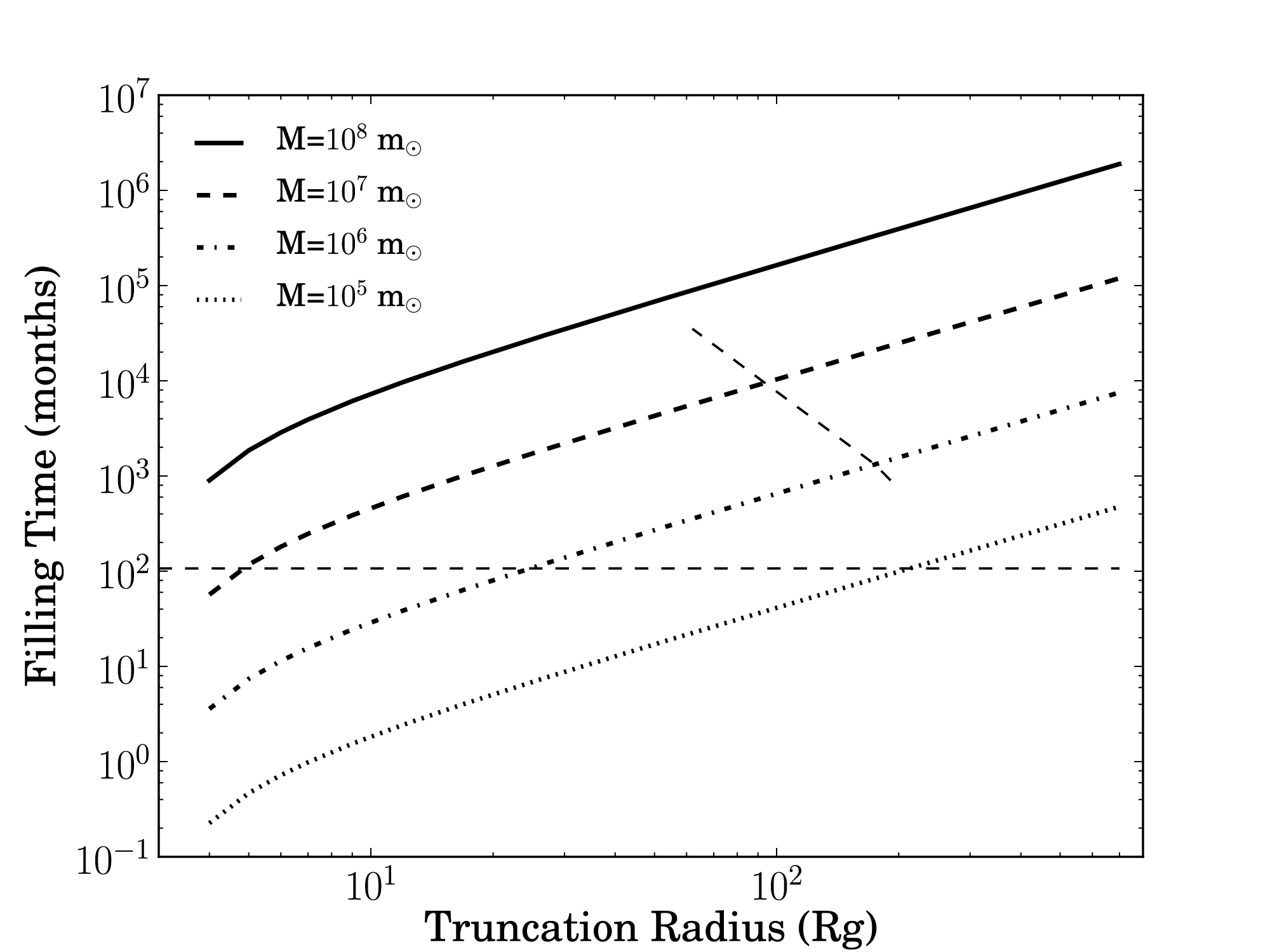}}
\caption[Time needed to fill an inner disc from a truncation radius]{\label{fig:timefill_a} Time needed to fill an inner accretion disc previously truncated at $R_{\rm{trunc}}$. This assumes an innermost stable orbit at 3R$_{\rm{g}}$ and Eddington-limited accretion. The dashed horizontal line shows the upper limit 
on the rise time for the flare in \ngc (107 months). The diagonal dashed 
line indicates the truncation radius where thermal emission from the inner disc 
is 130 times greater than that of the outer disc in the 0.2-2 keV band,
for the range of possible masses inferred for the nuclear black hole of \ngc. 
The time needed to produce the measured flux rise is always greater 
than the observed rise time for this range of black hole masses.}

\end{figure}

In Fig.~\ref{fig:timefill_a} we plot $\tau_{\rm{fill}}$ against the 
truncation radius for $M_{BH}$ ranging between $10^{5}$ and $10^{8}$
 M$_{\odot}$, assuming that the accretion rate is Eddington limited ($M_{\rm{edd}}=1$), $\alpha=0.1$ and $R_{0}=3R_{\rm{g}}$. 

We see that it is possible to fill the inner disc in $\leq107$ months for 
a limited range of $M_{\rm{BH}}$ and $R_{\rm{trunc}}$. To further constrain 
$R_{trunc}$ we estimate the size of the inner disc which will have 130
times more thermal 0.2--2 keV flux than the outer disc, i.e. which
would duplicate the change in flux between the 1993 \rosat 
and the 2002 \xmm slew observations. Here we used a Novikov-Thorne disc \citep{NovThorne} with colour corrections as described in \citet{Done12}.
These constraints for $7\times10^{5}<M_{BH}<4.6\times10^{7}$\msolarns, 
covering the range
estimated for the nuclear black hole of \ngcns, are shown in Fig.~\ref{fig:timefill_a}.
No solution is possible for $R_{\rm{trunc}}$ and $M_{\rm{BH}}$, within the
allowed mass range, which allows the flux
to rise sufficiently quickly if the emission comes from a thin disc. The possibility of an off-nuclear, lower
mass black hole, was discussed in section 3.2.
%Therefore we
%need to invoke an instability mechanism to boost the flux coming from the inner disc.
%In this way the truncation radius may be smaller and the timescales shorter.

%\cite{Belloni97a} and \cite{Belloni97b} have explained fast, soft,
%flares seen in some  Galactic binary systems in terms of a rapid disappearance and slower refilling of the central regions of the accretion disc. 
In the BHB, GRS~1915+105, soft flares are seen with a
rise and decay time of seconds, an order of magnitude too fast to be explained
by a complete filling and emptying of an inner disc on the viscous time scale, even though spectrally this
is a very attractive solution \citep{Belloni97a}. To explain these very fast spectral/flux variations, the Lightman-Eardley (LE) disc instability has been 
invoked \citep{Lightman1974,Cannizzo96,Belloni97a,Belloni97b}. 
In this scenario, each flare (preceded by a quiescent phase\footnote{Note that the quiescent state in X-ray binaries is different from the quiescent phase, which is normally seen in black hole binary transients: the former is a low-luminosity active state, while the latter is a quiet phase where the X-ray luminosity is negligible. }) constitutes a cycle with different phases. 

\begin{itemize}

\item Quiescent phase: during this phase the disc is truncated at a certain radius, with the central region either empty or filled with gas whose radiation is too soft to be detected. Slowly, the disc is refilled by a steady accretion rate $\dot{M_{0}}$ from the outer regions. 
%Each annulus of the disc  moves forward at the viscous time scale. During this phase no changes are seen in the truncation radius.

\item Rise phase: at some point the radiation pressure in the inner disc exceeds the gas pressure.
%at some point one of the annuli will reach the so-called unstable point (\cite{Lightman1974}) where the accretion rate is higher than the outer accretion rate $\dot{M_{0}}$. 
This causes a chain reaction (due to the LE disc instability) that switches on the inner disc (that has been now refilled and has the form of a slim hot disc). During this phase the flux from the source rises quickly and becomes much softer. 

\item Outburst phase (\textit{plateau}): now the disc extends down to an orbit that is very close to or coincident with the innermost stable circular orbit (ISCO) and a smaller hot radius can be observed. During this phase the emission is dominated by the thermal contribution from the disc, that is now hot and bright. The time-scale governing this phase is unknown, even though \citet{Belloni97b} noticed that the duration of the outburst phase correlates with the duration of the quiescent phase. 

\item Decay phase: eventually the inner disc runs out of fuel (because it is accreted onto the black hole faster than it is replenished) and switches off, either cooling and jumping back to an accretion rate smaller than $\dot{M_{0}}$ or emptying completely, leaving a new hole in the disc. The flux will therefore decay very quickly, bringing the source back to the quiescent phase.
\end{itemize}

%The rise time $\tau_{rise}$ (rise phase) is governed by the LE disc instability. 

The temporal development of the LE instability is compatible with that of the flare seen
in \ngc. The instability begins when the radiation pressure becomes comparable with the gas pressure. A heating wave is generated at the inner edge of the disc which propagates backwards through the disc at the sound speed ($c_{\rm{s}}$). The heating increases the local viscosity, scale height of the disc and 
the local accretion rate \citep{Cannizzo96}. At this point the disc is no longer a thin
Shakura-Sunyaev disc but still emits as a thermalised plasma. 
The time taken to fully heat the inner disc gives the rise time of the flare:

\begin{equation}
    \tau _{\rm{rise}} \gtrsim R_{\rm{trunc}} / c_{\rm{s}} \sim 1.5\times10^{4} \big( \frac{R_{\rm{trunc}}}{R_{\rm{g}}} \big) M_{6}\,\,\mathrm{seconds} 
\end{equation}

about a month for $R_{\rm{trunc}} = 100~R_{\rm{g}}$.

When the disc is emptied, as the hot radiating matter is accreted into
the black hole
more quickly than it is replenished  \citep{Belloni97a},  the flare 
decays (decay phase). The time $ \tau_{\rm{decay}}$ required to empty the inner disc (or to let it go back to a  tenuous plasma regime) cannot be shorter than the dynamical timescale, therefore it is given by:

\begin{equation}
    \tau_{\rm{decay}} \gtrsim 6 M_{6} \big(\frac{R_{\rm{trunc}}}{R_{\rm{g}}}\big)^{3/2}\mathrm{seconds}
\end{equation}

a few hours for $M_{\rm{BH}} \sim 10^{6}M_{\odot}$ and $R_{\rm{trunc}} = 100R_{\rm{g}}$. 
We note here that our estimates are stringent lower limits, based on the assumption that the accretion rate is Eddington limited. However, it is important to notice that, given the timescales discussed above, the rise time being longer than the decay time is 
a necessary condition for our interpretation to hold. Our results show
that the rise time ($\lesssim$107 months) is consistent with being longer than 
the decay time ($\sim$36 months), therefore this constraint is not violated. 

As the inner disc is emptied and refilled, the flares should repeat on a timescale which
is governed by the filling or viscous time at the truncation radius, i.e. decades for $M_{\rm{BH}}\sim10^{6} M_{\odot}$. A similar large flare in the Sy 1.9 galaxy IC~3599, which may have been 
produced by the same mechanism, repeated after 20 years \citep{Grupe15}.

Numerical simulations by \citet{xue11} show that the magnitude of the flare produced by
an instability in the standard disc model is strongly influenced by the BH spin, while
the timescale primarily depends on $M_{\rm{BH}}$ and viscosity. Their simulations found a short flare with 
a duty cycle $\lesssim3$\% which would imply a long ($\geq50$ year) quiescent phase for \ngcns.
Nevertheless, analytical work by \citet{nayakshin00} and \citet{janczern05}, which include
the effects of a comptonisation zone, jet and non-standard viscosity, and produce longer flares compatible with those seen in GRS~1915+105, would allow a shorter repeat time.

\section{Summary}

Based on new data we have shown that it is difficult to reconcile the
rise time of the soft X-ray flare seen in NGC 3599 with its previous classification 
as a classical, fast-rising, short plateau, tidal disruption event. 
It does, however, fit into the emerging 
scheme of a disruption which has led to a late, distant, circularisation where 
the rise is flatter and the peak lasts longer or alternatively to the 
tidal stripping of a red giant star.

If the factor 100 soft X-ray flare is interpreted as a change in the accretion 
from a persistent AGN, we find that the rise time of $\leq 9$ years is too short to be explained purely by the bulk motion of disc material about a $10^{6} M_{\odot}$ BH.
The observed timescales are compatible with the Lightman-Eardley instability which boosts the thermal soft X-ray emission by heating the inner disk and raising its scale height. 
This mechanism predicts that flares will repeat on
the viscous timescale of the truncation radius, i.e. every few decades.
The behaviour of NGC 3599 would then be analogous to flares seen in certain Galactic binary systems, but is too rapid to be equivalent to the state changes which are seen
in those systems.

\section*{Acknowledgments}

We thank the anonymous referee for useful comments which 
improved the paper.
RDS and SK would like to thank Giuseppe Lodato, Elena Rossi and the European
Science Foundation for hosting the 
Tidal Disruption conference in Favignana.
RDS would also like to acknowledge Giovanni Miniutti and Margarit Giustini 
for interesting and helpful discussions and Pilar Esquej for providing the optical spectrum.
SEM acknowledges the ESA research fellowship program and the Violette and Samuel Glasstone foundation.


\begin{thebibliography}{68}

%\bibitem[Alexander 2012]{Alexander} Alexander, T. 2012, EPJWC, 3905001A
\bibitem[\protect\citeauthoryear{Arcavi et~al.}{2014}]{Arcavi14} Arcavi, I., Gal-Yam, A., Sullivan, M., Pan, Y-C., Cenko, S. et al., 2014, ApJ, 793,38
\bibitem[\protect\citeauthoryear{Bade, Komossa \& Dahlem}{1996}]{Bade96} Bade, N., Komossa, S., Dahlem, M. 1996, A\&A, 309, L35 
%\bibitem[Begelman \& Armitage 2014]{Begel14} Begelman, M. \& Armitage, P. 2014 ApJ, 782, 18
\bibitem[\protect\citeauthoryear{Belloni et~al.}{1997a}]{Belloni97a}] Belloni, T., Mendez, M., King, A., van der Klis, M., van Paradijs, J., 1997, ApJ, 479, L145
\bibitem[\protect\citeauthoryear{Belloni et~al.}{1997b}]{Belloni97b}] Belloni, T., Mendez, M., King, A., van der Klis, M., van Paradijs, J.,  1997, ApJ, 488, 109
%\bibitem[Belloni 2010]{Belloni10} Belloni, T. 2010, LNP 794, 53 
%\bibitem[Belloni 2012]{Belloni12} Belloni, T. 2012 
%\bibitem[Bentz et~al. 2009]{Bentz} Bentz, M., Peterson, B., Pogge, R., \& Vestergaard, M. 2009, ApJ, 694, L166
%\bibitem[Bogdanovic et~al. 2004]{Bogdanovic} Bogdanovic, T, Eracleous, M., Mahadevan, S., Sigurdsson, S., Laguna, P. 2004, ApJ, 610, 707 
%\bibitem[Boller et~al. 1997]{Boller07} Boller, Th.; Brandt, W., Fabian, A. C., Fink, H. 2007, MNRAS, 289, 393 
%\bibitem[Brandt, Pounds \& Fink 1995]{Brandt} Brandt, W., Pounds, K. \& Fink, H., 1995, MNRAS 273, L47. 
\bibitem[\protect\citeauthoryear{Bloom et~al.}{2011}]{Bloom} Bloom, J et al., 2011, Sci, 333, 203 
%\bibitem[\protect\citeauthoryear{Blustin et~al.}{2002}]{Blustin} Blustin, A., Branduardi-Raymont, G., Behar, E., Kaastra, J. et al., 2002, A\&A, 392, 453
%\bibitem[Bruzual \& Charlot 2003]{Bruzual} Bruzual, G. \& Charlot, S. 2003, MNRAS, 344, 1000
%\bibitem[Burrows et~al. 2005]{Burrows05} Burrows, D., Hill, J., Nousek, J. et al. 2005, Space Sci. Rev., 120, 165  
\bibitem[\protect\citeauthoryear{Burrows et~al.}{2011}]{Burrows11} Burrows, D., Kennea, J., Ghisellini, G. et al., 2011, Nat., 476, 421
%\bibitem[Caldwell, Rose \& Concannon 2003]{Caldwell} Caldwell, N., Rose, J., \& Concannon, K. 2003, ApJ, 125, 2891
\bibitem[\protect\citeauthoryear{Caldwell et~al.}{2003}]{Caldwell} Caldwell, N., Rose, J., \& Concannon, K., 2003, ApJ, 125, 2891
%\bibitem[Cannizzo, Lee \& Goodman 1990]{Cannizzo} Cannizzo, J., Lee, H. \& Goodman, J. 1990, ApJ, 351, 38 
\bibitem[\protect\citeauthoryear{Campana et~al.}{2015}]{Campana15} Campana, S., Mainetti, D., Colpi, M., Lodato, G, D'Avanzo, P., Evans, P.A., \& Moretti, A., 2015, A\&A submitted
\bibitem[\protect\citeauthoryear{Cannizzo}{1996}]{Cannizzo96} Cannizzo, J., 1996, ApJ, 466, L31 
\bibitem[\protect\citeauthoryear{Cappelluti et~al.}{2009}]{Cappelluti09} Cappelluti, N., Ajello, M., Rebusco, P. et al., 2009, A\&A, 495, L9
\bibitem[\protect\citeauthoryear{Caramete \& Biermann}{2010}]{Caramete10} Caramete, L., Biermann, P., 2010, A\&A, 521, 55
%\bibitem[Cash 1979]{Cash} Cash, W. 1979, ApJ 228, 939
\bibitem[\protect\citeauthoryear{Cenko et~al.}{2012a}]{Cenko} Cenko, S., Bloom, J., Kulkarni, S. et al., 2012, MNRAS, 420, 2684
\bibitem[\protect\citeauthoryear{Cenko et~al.}{2012b}]{Cenko12b} Cenko, S., Krimm, H., Horesh, A. et al., 2012, ApJ, 753, 77
%\bibitem[Chen et~al. 2009]{Chen} Chen, X., Madau, P., Sesana, A., Liu, F., 2009, ApJ, 697, L149 
\bibitem[\protect\citeauthoryear{Davis et~al.}{2011}]{Davis11} Davis, S., Narayan, R., Zhu, Y. et al., 2011, ApJ, 734, 111
\bibitem[\protect\citeauthoryear{Drake et~al.}{2009}]{catalina} Drake, A.J. et al., 2009, ApJ, 696, 870
%\bibitem[Comastri et~al. 2002]{Comastri02} Comastri, A., Mignoli, M., Ciliegi, P. et al. 2002, ApJ, 571, 771
%\bibitem[Done et~al. 2012]{Done12} Done, C., Davis, S., Jin, C., Blaes, O., Ward, M. 2012, MNRAS 420, 1848
%\bibitem[\protect\citeauthoryear{Donley et~al.}{2002}]{Donley} Donley, J., Brandt, W., Eracleous, M., Boller, Th. 2002, AJ 124, 1308
%\bibitem[Elitzur \& Shlosman 2006]{Elitzur06} Elitzur, M. \& Shlosman, I., 2006, ApJ, 648, L101
%\bibitem[Done et~al. 2012]{Done12} Done, C, Davis, S, Jin, C., Blaes, O., Ward, M. 2012, 2012, MNRAS 420, 1848
\bibitem[\protect\citeauthoryear{Done et~al.}{2012}]{Done12} Done, C, Davis, S, Jin, C., Blaes, O., Ward, M., 2012, MNRAS 420, 1848
\bibitem[\protect\citeauthoryear{Esin et~al.}{1997}]{Esin97} Esin, A., McClintock, J., Narayan, R., 1997, ApJ, 489, 865
\bibitem[\protect\citeauthoryear{Esquej et~al.}{2007}]{Esquej07} Esquej, P., Saxton, R., Freyberg, M. et al., 2007, A\&A, 462L, 49
\bibitem[\protect\citeauthoryear{Esquej et~al.}{2008}]{Esquej08} Esquej, P., Saxton, R., Komossa, S., Read, A., Freyberg, M. J., 2008, A\&A, 489, 543 
\bibitem[\protect\citeauthoryear{Esquej et~al.}{2012}]{Esquej12} Esquej, P., Saxton, R., Komossa, S., Read, A., 2012, EPJWC, 3902004E
\bibitem[\protect\citeauthoryear{Farrell et~al.}{2009}]{HLX1} Farrell, S., Webb, N., Barret, D., Godet, O., Rodrigues, J., 2009, Nat., 460, 73 
%\bibitem[Evans et~al 2009]{Evans} Evans, P., Beardmore, A., Page, K. et al. 2009, MNRAS, 397, 1177 
%\bibitem[Fabian et~al. 2011]{Fabian11} Fabian, A., et al. 2011, arXiv1108.5988, MNRAS, in press 
%\bibitem[Fossati 2000]{Fossati} Fossati, G., Celotti, A., Chiaberge, M. et al. 2000, ApJ, 541, 166 
\bibitem[\protect\citeauthoryear{Frank, King \& Rayne}{1992}]{FKR} Frank, J., King, A. \& Rayne, D. 1992, Accretion Power in Astrophysics (Cambridge, Cambridge Univ. Press)
\bibitem[\protect\citeauthoryear{Gabriel et~al.}{2003}]{Gabriel} Gabriel, C. et al. : \emph{}, In ASP Conf. Ser., Vol. 314, ADASS Xiii ed.  Oschenbein, F., Allen, M. \& Egret, D., 759 (2003).
%\bibitem[Gezari et~al. 2003]{Gezari03} Gezari, S., Halpern, J., Komossa, S., Grupe, D., Leighly, K. 2003, ApJ, 592, 42
\bibitem[\protect\citeauthoryear{Gezari et~al.}{2006}]{Gezari06} Gezari, S., Martin, D., Milliard, B. et al., 2006, ApJ, 653L, 25 
\bibitem[\protect\citeauthoryear{Gezari et~al.}{2008}]{Gezari08} Gezari, S., Basa, S., Martin, D. et al., 2008, ApJ, 676, 944
\bibitem[\protect\citeauthoryear{Gezari et~al.}{2009}]{Gezari09} Gezari, S., Heckman, T., Cenko, S. et al., 2009, ApJ, 698, 1367 
\bibitem[\protect\citeauthoryear{Gezari et~al.}{2012}]{Gezari12} Gezari, S., Chornock, R., Rest, A et al., 2012, Nat., 485, 217
%\bibitem[Goad et~al. 2007]{Goad} Goad, M., Tyler, L., Beardmore, A. et al. 2007, A\&A, 476, 1401
%\bibitem[Green \& Ho 2005]{greenHo2005} Green, J. \& Ho, L. 2005, ApJ, 630, 122
\bibitem[\protect\citeauthoryear{Greiner et~al.}{2000}]{Greiner} Greiner, J., Schwarz, R., Zharikov, S., Orio, M., 2000, A\&A, 362, L25 
%\bibitem[\protect\citeauthoryear{Grupe et~al.}{1995}]{Grupe95WPVS} Grupe, D., Beuerman, K., Mannheim, K. et al., 1995, A\&A, 300, L21
%\bibitem[\protect\citeauthoryear{Grupe et~al.}{1995b}]{Grupe95IC3599} Grupe, D., Beuerman, K., Mannheim, K. et al., 1995, A\&A, 299, L51
\bibitem[\protect\citeauthoryear{Grupe et~al.}{1999}]{Grupe99} Grupe, D., Thomas, H.-C., Leighly, K., 1999, A\&A 350, L31
%\bibitem[Grupe, Komossa \& Gallo 2007]{Grupe07} Grupe, D.,  Komossa, S. \& Gallo, L.2007, ApJ, 668, L111 
%\bibitem[Grupe et~al. 2008]{Grupe08} Grupe, D., Leighly, K.~M., Komossa, S. 2008, AJ, 136, 234
\bibitem[\protect\citeauthoryear{Grupe, Komossa \& Saxton}{2015}]{Grupe15} Grupe, D., Komossa, S., Saxton, R., 2015, ApJ, 803, L28
%\bibitem[Guillochon, Manukian \& Ramirez-Ruiz 2013]{GuillochonPSJ} Guillochon J., Manukian H., Ramirez-Ruiz E., 2013, ApJ, preprint (arXiv:1304.6397)
\bibitem[\protect\citeauthoryear{Guillochon \& Ramirez-Ruiz}{2015}]{Guill15} Guillochon J., Ramirez-Ruiz E., 2015, ApJ, preprint (arXiv:1501.05306)
%\bibitem[Guillochon \& Ramirez-Ruiz 2015]{Guill15} Guillochon J. \&  Ramirez-Ruiz E., 2015, ApJ, preprint (arXiv:1501.05306)
\bibitem[\protect\citeauthoryear{Halpern, Gezari \& Komossa}{2004}]{Halpern04} Halpern, J., Gezari, S., Komossa, S., 2004, ApJ, 604, 572
\bibitem[\protect\citeauthoryear{Hayasaki, Stone \& Loeb}{2015}]{Hyak15} Hayasaki, K., Stone, N., Loeb, A., 2015, arXiv:1501.05207
\bibitem[\protect\citeauthoryear{Hills}{1975}]{Hills} Hills, J., 1975, Nat. 254, 295
%\bibitem[Ivanov, Polnarev \& Saha 2005]{Ivanov05} Ivanov, P., Polnarev, A.\& Saha, P., 2005, MNRAS, 358, 1361
%\bibitem[Jansen et~al., 2001]{jansen} Jansen, F. et~al, 2001. \aap, 365, L1-6
%Giannios, D \& Metzger, B 2011, MNRAS (arXiv:1102.1429) \\
\bibitem[\protect\citeauthoryear{Janiuk \& Czerny}{2005}]{janczern05}Janiuk A., Czerny B., 2005, MNRAS, 356, 205
\bibitem[\protect\citeauthoryear{Kalberla et~al.}{2005}]{Kalberla} Kalberla, P., Burton, W., Hartmann, D.
 et al., 2005, A\&A, 440, 775
%\bibitem[Kauffmann et~al. 2003]{Kauffmann03}  Kauffmann, G., Heckman, T., Tremonti, C. et al., 2003, MNRAS, 346, 1055
\bibitem[\protect\citeauthoryear{Kauffmann \& Heckman}{2004}]{Kauffmann04}  Kauffmann, G., Heckman, T., 2004, A\&A 405, 5
\bibitem[\protect\citeauthoryear{Karas \& Subr}{2007}]{KarasSubr07} Karas, V., Subr, L., 2007, A\&A, 470, 11 
\bibitem[\protect\citeauthoryear{Khabibullin \& Sazonov}{2014}]{KhabSaz14} Khabibullin, I, Sazonov, S. 2014, MNRAS, 441, 1041
%\bibitem[Khokhlov \& Melia 1996]{Khokhlov} Khokhlov, A. \& Melia, F. 1996, ApJ 457, L61 
%\bibitem[Koerding, Jester \& Fender 2006]{Kording06} Kording, E., Jester, S. \&Fender, R. 2006, MNRAS, 372, 1366
%\bibitem[\protect\citeauthoryear{Komossa}{2002}]{Komossa02} Komossa, S., 2002, RvMA 15, 27
\bibitem[\protect\citeauthoryear{Komossa \& Bade}{1999}]{KomossaBade} Komossa, S., Bade, N., 1999, A\&A, 343, 775 
%\bibitem[\protect\citeauthoryear{Komossa \& Fink}{1997}]{KomossaFink97} Komossa S., Fink H., 1997, In: Meyer-Hofmeister E., Spruit H. (eds.) Accretion Disks – New Aspects. Lecture Notes in Physics 487, A\&A, 343, 775 
\bibitem[\protect\citeauthoryear{Komossa \& Greiner}{1999}]{Komossa99b} Komossa, S., Greiner, J., 1999, A\&A, 349, L45 
\bibitem[\protect\citeauthoryear{Komossa et~al.}{2004}]{Komossa1242} Komossa, S., Halpern, J., Schartel, N. et al. 2004, ApJ, 603, L17 
\bibitem[\protect\citeauthoryear{Komossa}{2005}]{Komossa05} Komossa, S. 2005, in: Growing Black Holes: accretion in a cosmological context, 159
\bibitem[\protect\citeauthoryear{Komossa et~al.}{2008}]{Komossa08} Komossa, S., Zhou, H., Wang, T. et al., 2008, ApJ, 678, 13
%\bibitem[Komossa \& Merritt 2008]{KomossaMerritt08} Komossa, S., \& Merritt, D., 2008, ApJ, 683, L21 
%\bibitem[Komossa et~al. 2009]{Komossa09} Komossa, S., Zhou, H., Rau, A. et al. 2009, ApJ, 701, 105
\bibitem[\protect\citeauthoryear{Lamastra et~al.}{2009}]{Lamastra} Lamastra, A., Bianchi, S., Matt, G. et al., 2009, A\&A, 504, 73
%\bibitem[Lamer, Uttley, McHardy 2003]{Lamer03} Lamer, G., Uttley, P., McHardy, I. M. 2003, MNRAS, 342, L41 
\bibitem[\protect\citeauthoryear{Lauer et~al.}{2007}]{Lauer07} Lauer, T., Faber, S., Richstone, D., et al., 2007, ApJ, 662, 808
%\bibitem[Lehto \& Valtonen 1996]{LehtoValt} Lehto, H. \& Valtonen, M. 1996, ApJ, 460, L207 
%\bibitem[Leighly et~al. 2009]{Leighly09} Leighly, K., Hamann, F., Casebeer, D. \& Grupe, D.2009, ApJ, 701, 176 
%\bibitem[Levan et~al. 2011]{Levan} Levan, A., Tanvir, N., Cenko, S. et al.  2011, Sci 333, 199
%\bibitem[\protect\citeauthoryear{Li, Narayan \& Menou}{2002}]{Li02} Li, L-X., Narayan, R., Menou, K., 2002, ApJ, 576, 753	
\bibitem[\protect\citeauthoryear{Lightman \& Eardley}{1974}]{Lightman1974} Lightman, A. P., Eardley, D. M., 1974, ApJ, 187, L1
\bibitem[\protect\citeauthoryear{Lin et~al.}{2011}]{Lin11} Lin, D. Carrasco, R., Grupe, D., Webb, N., Barret, D. et al. 2011, ApJ, 738, 52
%\bibitem[\protect\citeauthoryear{Liu, Li \& Chen}{2009}]{LLC09}Liu, F., Li, S., Chen, X., 2009,ApJ, 709, 133
%\bibitem[\protect\citeauthoryear{Liu, Li \& Komossa}{2014}]{LiuLiKom14}Liu, F., Li, S.,  Komossa, K. 2014, ApJ, 786, 103
\bibitem[\protect\citeauthoryear{Longinotti et~al.}{2013}]{Longinotti13}Longinotti, A., Krongold, Y., Kriss, G., Ely, J., Gallo, L., Grupe, D. et al., 2013, ApJ, 766, 104 

%\bibitem[Lodato, King \& Pringle 2009]{Lodato09} Lodato, G., King, A.R. \& Pringle, J.E.  2009, MNRAS, 392, 332 
\bibitem[\protect\citeauthoryear{Luminet}{1985}]{Luminet} Luminet, J.-P., 1985, AnPh, 10, 101

\bibitem[\protect\citeauthoryear{MacLeod, Guillochon \& Ramirez-Ruiz}{2012}]{MacLeod12} MacLeod, M., Guillichon, J., Ramirez-Ruiz, E., 2012, ApJ, 757, 134.
%\bibitem[Magorrian \& Tremaine 1999]{Magorrian} Magorrian, J. \& Tremaine, S. 
%1999, MNRAS 309, 447
\bibitem[\protect\citeauthoryear{Maksym et~al.}{2010}]{Maksym10} Maksym, W., Ulmer, M., Eracleous, M.,  2010, ApJ, 722, 1035
\bibitem[\protect\citeauthoryear{Maksym, Lin \& Irwin}{2014}]{Maksym14} Maksym, W., Lin, D., Irwin, J. 2014, ApJ, 792, 29
%\bibitem[Marconi \& Hunt 2003]{MarconiHunt} Marconi, A., Hunt, L. 2003, ApJ, 589, L21
%\bibitem[Magnier 2007]{PanS} Magnier E. 2007, ASPC, 364, 153, ed. C. Sterken (San Francisco: ASP)
%\bibitem[Matt, Guainazzi \& Maiolino 2003]{Matt03}Matt, G., Guainazzi, M. \& Maiolino, R. 2003, MNRAS, 342, 422 
%\bibitem[McHardy et~al. 2006]{McHardy06} McHardy, I., Koerding, E.; Knigge, C.; Uttley, P.; Fender, R. P 2006, Nature, 444, 730 
%\bibitem[Milosavljevic, Merritt \& Ho 2006]{Milosavljevic} Milosavljevic, M., Merritt, D. \& Ho, L. 2006, ApJ 652, 120
\bibitem[\protect\citeauthoryear{Merloni et~al.}{2015}]{Merloni} Merloni, A., Dwelly, T., Salvato, M., Georgakakis, A., Greiner, J., 2015, arXiv:1503.04870
%\bibitem[Motta et~al. 2012]{Motta12} Motta, S., Homan, J., Munoz-Darias, T. et al. 2012, MNRAS 427, 595
%\bibitem[Miniutti \& Fabian 2004]{minfab04} Miniutti, G. \& Fabian, A. 2004, MNRAS, 349, 1435
%\bibitem[Miniutti et~al. 2009]{Miniutti09} Miniutti, G., Fabian, A., Brandt, W., Gallo, L., Boller, Th., 2009, MNRAS, 396, 85.
%\bibitem[Miniutti et~al. 2012]{Miniutti12} Miniutti, G., Brandt, W., Schneider, P., Fabian, A., Gallo, L. \&  Boller, Th. 2012, MNRAS, 425, 1718
%\bibitem[Miniutti et~al. 2013]{Miniutti13} Miniutti, G., Saxton, R. Rodrıguez–Pascual, P., Read, A., Esquej, P. et~al. 2013, MNRAS, 433, 1764
\bibitem[\protect\citeauthoryear{Nayakshin, Rappaport \& Melia}{2000}]{nayakshin00}Nayakshin S., Rappaport S., Melia F., 2000, ApJ, 535, 798
%\bibitem[Nagao, Taniguchi \& Murayama 2000]{Nagao} Nagao, T., Taniguchi, Y. \& Murayama, T. 2000, AJ, 119, 2605
%\bibitem[Narayan \& Yi 1994]{Narayan} Narayan, R. \& Yi, I. 1994, ApJ, 428, L13 
%\bibitem[Nicastro 2000]{Nicastro} Nicastro, F. 2000, ApJ, 530, L65  
\bibitem[\protect\citeauthoryear{Nikolajuk \& Walter}{2013}]{Walter13} Nikolajuk, M.,  Walter, R., 2013, A\&A, 552, 75
%\bibitem[Nixon \& Salvesen 2014]{Nixon14} Nixon, C. \& Salvesen G. 2014, MNRAS 437,3994
\bibitem[\protect\citeauthoryear{Novikov \& Thorne}{1973}]{NovThorne} Novikov I. D., Thorne K. S., 1973, in Black holes (Les astres occlus), 343–450
%\bibitem[Otani, et~al. 1996]{Otani} Otani, C., Kii, T., Miya, K. 1996, rftu.proc, 491O
\bibitem[\protect\citeauthoryear{Pasham et~al.}{2015}]{Pasham} Pasham, D., Cenko, S., Levan, A., Bower, G., Horesh, A. et al., 2015, arXiv:1502.01345
\bibitem[\protect\citeauthoryear{Phinney}{1989}]{Phinney} Phinney, E.S., 1989, vol 136 of IAU symposium , 543
\bibitem[\protect\citeauthoryear{Piran et~al.}{2015}]{Piran15} Piran, T., Svirski, G., Krolik, J., Cheng, R., Shiokawa, H., 2015, arXiv:1502.05792 
%\bibitem[Poole et~al. 2008]{Poole} Poole, T. et al. 2008, MNRAS, 383, 627
\bibitem[\protect\citeauthoryear{Predehl et~al.}{2010}]{eRosita} Predehl, P., Boehringer, H., Brunner, H. et al. 2010, SPIE, 7732, 23
%\bibitem[Ramirez-Ruiz \& Rosswog 2009]{Rosswog} Ramirez-Ruiz, E., \& Rosswog, S. 2009, ApJL, 697, L77
%\bibitem[Read et~al. 2008]{Read08} Read, A., Saxton, R., Torres, M. et al.  2008, A\&A, 482, L1
%\bibitem[Rau et~al. 2009]{PTF} Rau, A., Kulkarni, S., Law, N., Bloom, J., Ciardi, D. et~al. 2009, PASP, 121, 1334
\bibitem[\protect\citeauthoryear{Rees}{1988}]{Rees88} Rees, 1988, Nature, 333, 523 
%\bibitem[Risaliti et~al. 2005]{Risaliti05} Risaliti, G., Elvis, M., Fabbiano, G., Baldi, A., Zezas, A. 2005, ApJ, 623, L93 
%\bibitem[Risaliti et~al. 2009]{Risaliti09} Risaliti, G., Miniutti, G., Elvis, M. et al. 2009, ApJ, 696, 160
%\bibitem[Roming et~al. 2005]{Roming} Roming, P. et al. 2005, SSRv 120, 95
%\bibitem[Sault, Teuben \& Wright 1995]{Sault} Sault, R., Teuben, P. \& 
%Wright, M. 1995, ASPC, 77, 433
%\bibitem[Saxton et~al. 2008]{Saxton08} Saxton, R., Read, A., Esquej, P. et al.  2008, A\&A 480, 611 
%\bibitem[Saxton et~al., 2011]{Saxton11} Saxton, R., Read, A., Esquej, P., Miniutti, G., Alvarez, E. 2011, arXiv1106.3507S 
%\bibitem[Saxton et~al. 2011]{Saxton11} Saxton, R., Read, A., Esquej, P., Miniutti, G., Alvarez, E. 2011, In: "Narrow-Line Seyfert 1 Galaxies and Their Place in the Universe", Proceedings of Science, NLS1, 008
\bibitem[\protect\citeauthoryear{Saxton et~al.}{2012}]{Saxton12} Saxton, R., Read, A., Esquej, P. et al.,  2012, A\&A, 541, 106
\bibitem[\protect\citeauthoryear{Servillat et~al.}{2011}]{Servillat11} Servillat, M., Farrell, S., Lin, D. et al., 2011, ApJ, 743, 6
%\bibitem[Saxton et~al. 2014]{Saxton14} Saxton, R., Read, A., Komossa, S., Rodriguez-Pascual, P., Miniutti, G. et al. 2014, A\&A, 572, 1
\bibitem[\protect\citeauthoryear{Shakura \& Sunyaev}{1973}]{ShakSun73} Shakura, N., Sunyaev, R., 1973, A\&A, 24, 337 
%\bibitem[Shappee et~al. 2013]{Shappee} Shappee, B., Prieto, J., Grupe, D. et al. 2013, arXiv:1310.2241
\bibitem[\protect\citeauthoryear{Shiokawa et~al.}{2015}]{Shiok15} Shiokawa, H., Krolik, J., Cheng, R., Piran, T., Noble, S. 2015, arXiv:1501.04365 
%\bibitem[\protect\citeauthoryear{Strubbe \& Quataert}{2009}]{Strubbe09} Strubbe, L., Quataert, E., 2009, MNRAS, 400, 2070 
%\bibitem[Strubbe \& Quataert 2011]{Strubbe11} Strubbe, L.E. \& Quataert, E. 2011, MNRAS, 415, 168
%\bibitem[Takahashi 1996]{Takahashi} Takahashi, T., et al. 1996, ApJL, 470, L89 
%\bibitem[Trump et~al. 2009]{Trump09} Trump, J. et al. 2009, ApJ, 706, 797 
%\bibitem[\protect\citeauthoryear{Ulmer}{1999}]{Ulmer99} Ulmer, A. 1999, ApJ, 514, 180  
%\bibitem[Uttley \& McHardy 2004]{Uttley04} Uttley, P. \& McHardy, I. 2004, PThPS. 155, 170
\bibitem[\protect\citeauthoryear{Xue et~al.}{2011}]{xue11} Xue, L., Sadowski, A., Abramowicz, M., Lu, J.-F., 2011, ApJS, 195, 7
\bibitem[\protect\citeauthoryear{van Velzen et~al.}{2011}]{vanVelzen11} van Velzen, S. et al., 2011, ApJ, 741, 73
%van Velzen, S. et al. 2011b, MNRAS (arXiv:1104.4105) \\
%\bibitem[Kennedy, Miralda-Escude \& Kollmeier, 2010]{Kennedy10}  \\
%\bibitem[Vaughan, Edelson \& Warwick 2004]{Vaughan} Vaughan, S., Edelson, R. \& Warwick, R. 2004, MNRAS, 349, L1 
%\bibitem[Veilleux \& Osterbrook 1987]{veillandost} Veilleux, S., \& Osterbrock, D. E. 1987, ApJS, 63, 295
%\bibitem[Voges et~al. 1999]{Voges} Voges, W., Aschenbach, B., Boller, T.  et al. 1999, A\&A, 349, 389
\bibitem[\protect\citeauthoryear{Wang et~al.}{2011}]{Wang11} Wang, T.-G., Zhou, H.-Y., Wang, L.-F., Lu, H.-L., Xu, D., 2011, ApJ, 740, 85
%\bibitem[Wang \& Merritt 2004]{WangMerritt} Wang, J. \& Merritt, D. 2004, ApJ, 600, 149
%\bibitem[Yuan et~al. 2010]{Yuan} Yuan, W., Liu, B.~F., Zhou, H. \& Wang, T. 2010, ApJ, 723, 508  
\end{thebibliography}
\end{document}